\documentclass[%
 twocolumn,aip,
 sd,%
 amsmath,amssymb,
 reprint,%
]{revtex4-1}
\usepackage{siunitx}
\usepackage{graphicx}
\usepackage{dcolumn}
\usepackage{bm}
\begin{document}

\title{On the Angular Dependence of InP High Electron Mobility Transistors for Cryogenic Low Noise Amplifiers in a Magnetic Field}

\author{Isabel Harrysson Rodrigues}
\email{isabelr@chalmers.se}
\affiliation{GigaHertz Centre, Department of Microtechnology and Nanoscience - MC2, Chalmers University of Technology, SE-41296 Gothenburg, Sweden}
\author{David Niepce}
\affiliation{Quantum Technology Laboratory, Department of Microtechnology and Nanoscience - MC2, Chalmers University of Technology, SE-41296 Gothenburg, Sweden}
\author{Arsalan Pourkabirian}
\affiliation{Low Noise Factory AB, Nellickev\"agen 22, SE-41663 Gothenburg, Sweden}
\author{Giuseppe Moschetti}
\thanks{Present address: Qamcom Research and Technology AB, Falkenbergsgatan 3, SE-412 85 Gothenburg, Sweden}
\affiliation{Low Noise Factory AB, Nellickev\"agen 22, SE-41663 Gothenburg, Sweden}
\author{Joel Schleeh}
\affiliation{Low Noise Factory AB, Nellickev\"agen 22, SE-41663 Gothenburg, Sweden}
\author{Thilo Bauch}
\affiliation{Quantum Device Physics Laboratory, Department of Microtechnology and Nanoscience - MC2, Chalmers University of Technology, SE-41296 Gothenburg, Sweden}
\author{Jan Grahn}
\affiliation{GigaHertz Centre, Department of Microtechnology and Nanoscience - MC2, Chalmers University of Technology, SE-41296 Gothenburg, Sweden}


\begin{abstract}
The InGaAs-InAlAs-InP high electron mobility transistor (InP HEMT) is the preferred active device used in a cryogenic low noise amplifier (LNA) for sensitive detection of microwave signals.  We observed that an InP HEMT 0.3-14~GHz LNA at 2~K, where the in-going transistors were oriented perpendicular to a magnetic field, heavily degraded in gain and average noise temperature already up to 1.5~T. Dc measurements for InP HEMTs at 2~K revealed a strong reduction in the transistor output current as a function of static magnetic field up to 14~T. In contrast, the current reduction was insignificant when the InP HEMT was oriented parallel to the magnetic field. Given the transistor layout with large gate width/gate length ratio, the results suggest a strong geometrical magnetoresistance effect occurring in the InP HEMT. This was confirmed in the angular dependence of the transistor output current with respect to the magnetic field. Key device parameters such as transconductance and on-resistance were significantly affected at small angles and magnetic fields. The strong angular dependence of the InP HEMT output current in a magnetic field has important implications for the alignment of cryogenic LNAs in microwave detection experiments involving magnetic fields.
\end{abstract}

\maketitle 

In many sensitive detection systems, high electron mobility transistor (HEMT) low-noise amplifiers (LNAs) at cryogenic temperatures (1-10~K) are used to read out tiny microwave signals. Some of these systems rely on the presence of a strong magnetic field, {\it e.g.} in mass spectrometry \cite{mathur2008} or detection of dark matter. \cite{PhysRevD.42.1297}$^,$\cite{2018arXiv180100835B} A potential application for cryogenic LNAs in a magnetic field is magnetic resonance imaging.\cite{Johansen18} It has long been known that the sensitivity of the cryogenic LNA is affected by the presence of a magnetic field: when aligned perpendicular to the magnetic field, the noise temperature of a cryogenic AlGaAs-GaAs (GaAs) HEMT LNA was shown to be strongly degraded with increasing magnetic field.\cite{bradley} However, reports on the electrical behavior in a magnetic field for the cryogenic InGaAs-InAlAs-InP (InP) HEMT LNA - the standard component used in today's most sensitive microwave receivers - have so far been absent. Compared to previous work,\cite{bradley} we here report that the gain and noise properties for the cryogenic InP HEMT LNA are much more prone for degradation when exposed to a magnetic field. For the first time, we measured the InP HEMT at 2~K as a function of angular orientation with respect to the magnetic field. It is shown that the InP HEMT output current is limited by a strong geometrical magnetoresistance effect (gMR). The results suggest that even small misalignment of the cryogenic InP HEMT LNA in a magnetic environment is detrimental to read-out sensitivity.
\par
\begin{figure}[t]
    \centering
    \includegraphics[width=\columnwidth]{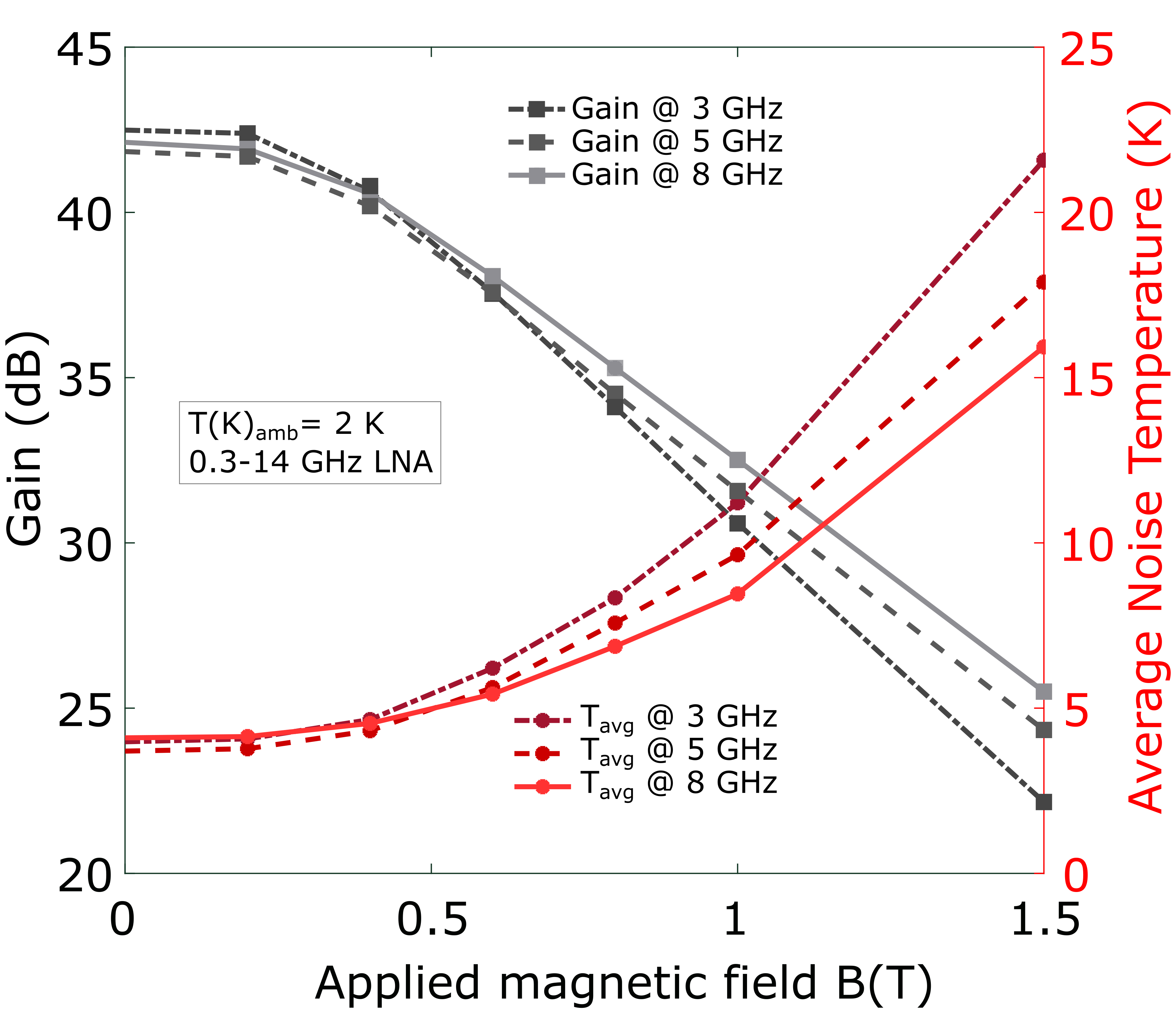}
    \caption{Gain and average noise temperature for the cryogenic InP HEMT MMIC LNA, measured at 2~K, for three different frequencies as a function of applied magnetic field. The LNA was aligned perpendicular to the magnetic field.}
    \label{fig:gain_noise_LNA_90deg}
\end{figure}

The sensitivity of the cryogenic InP HEMT LNA in a magnetic field was examined using a 10 T superconducting magnet. The LNA was a monolithic microwave integrated circuit (MMIC) chip consisting of three InP HEMT stages and passive components mounted in a gold-plated aluminum module. Neither the LNA module nor in-going passives exhibited any magnetic-field dependence as verified by additional experiments. The amplifier was a broadband design ranging from 0.3 to 14 GHz with a gain of 42 dB and average noise temperature of 4.2~K (0.06~dB).\cite{mmic_schleeh13TMTT} The LNA was mounted at the center of the magnet on the 2~K stage of an Oxford Instruments Triton 200 dilution refrigerator. In this arrangement, the LNA module was oriented perpendicular towards the magnetic field. Taking cable loss and an additional attenuator in account, the input signal was attenuated by 27~dB, which was necessary to thermalize the signal going from 300~K to 2~K as well as avoiding saturation of the amplifier. 
In Fig.~\ref{fig:gain_noise_LNA_90deg}, the gain and noise temperature for 3, 5 and 8 GHz are presented for the cryogenic InP HEMT LNA at 2 K when exposed to static magnetic fields up to 1.5~T. For all three measured frequencies, the LNA gain and noise temperature started to be affected around 0.25~T. From 0 to 1.5~T, the gain and average noise temperature degraded from 42 to 24~dB and 4 to 18~K, respectively, at 5 GHz. Beyond this field the gain was so low that we needed post amplification for measurements. The LNA degradation versus magnetic field was similar across the 3-8~GHz band. Compared to the earlier study for a GaAs HEMT LNA,\cite{bradley} the degradation for an InP HEMT LNA in the presence of a magnetic field is much stronger. This is connected to the higher electron mobility and sheet density of the two dimensional electron gas (2DEG) in the InP HEMT compared to the GaAs HEMT used in Ref.~\cite{bradley}.
\par
In order to better understand the cryogenic InP HEMT LNA behavior in a magnetic field, DC measurements were conducted on individual transistors at low temperature. The discrete InP HEMTs were fabricated in the same transistor technology\cite{hemt_schleeh12EDL} used for the cryogenic InP HEMT LNA measured in Fig.~\ref{fig:gain_noise_LNA_90deg}. We have fabricated two-finger device layouts with gate width $W_g$ ranging from 10 to 100 $\mu m$ and gate length $L_g$ from 60 to 250 $nm$. The DC-characterization was carried out in a Quantum Design Physical Property Measurement System (PPMS). The transistor was electrically connected through wire bonding and mounted in a matched LC-network on a sample holder in the vacuum chamber of the cryostat and cooled down to 2~K. A static magnetic field ranging up to 14~T was then applied and DC measurements were performed using a Keithley 2604B source meter.
\begin{figure}[t]
    \centering
    \includegraphics[width=\columnwidth]{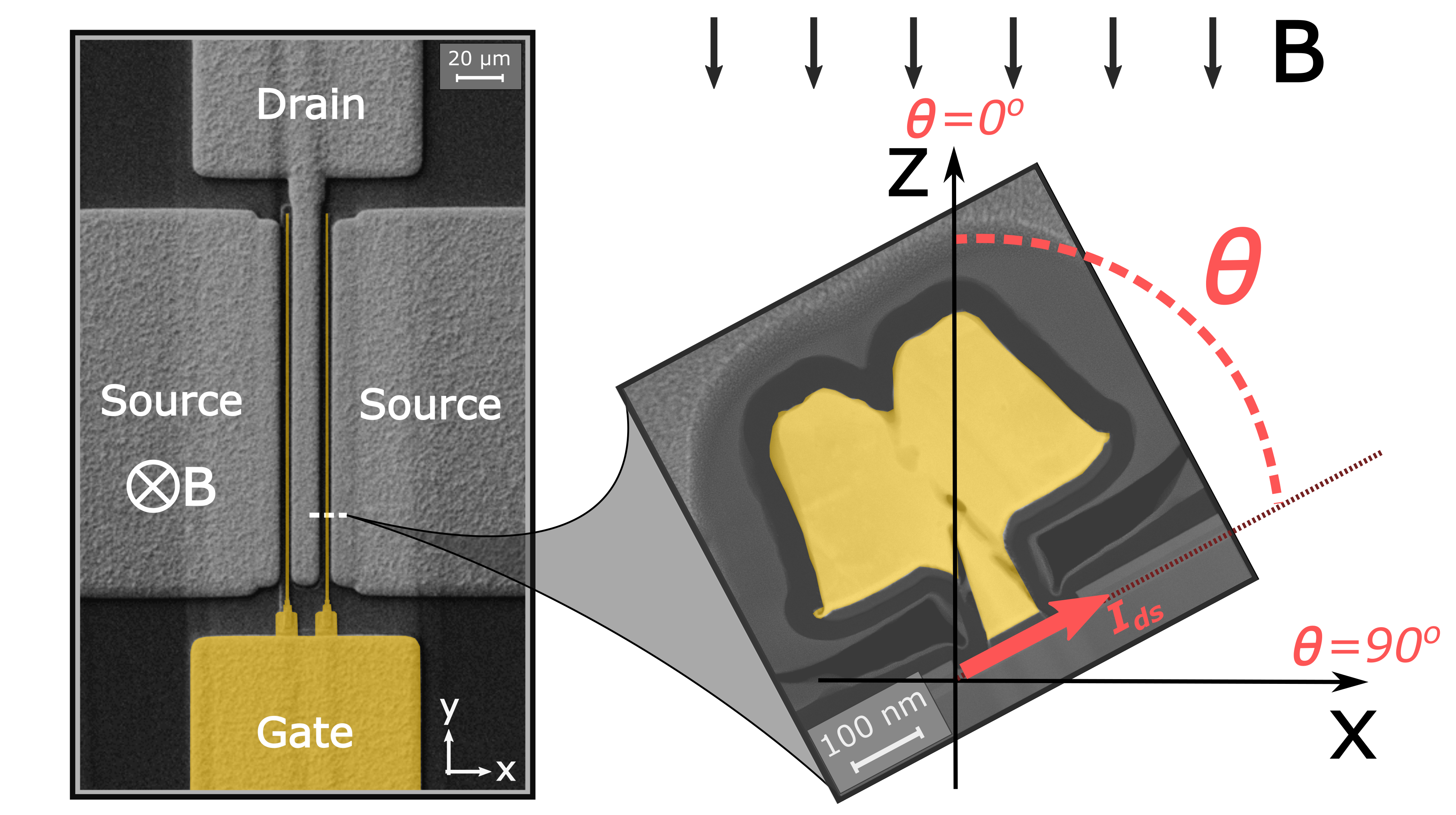}
    \caption{The two-finger InP HEMT shown in top view (by scanning electron microscopy) and side view (by cross-sectional transmission electron microscopy) perspective. The T-shaped gate is marked in yellow. The static magnetic field was applied in z-direction. The device was rotated from $\theta=0^\text{o}$ to $\theta=180^\text{o}$ corresponding to parallel and perpendicular orientation, respectively, of the HEMT output current with respect to the magnetic field.
    }
    \label{fig:cross_section}
\end{figure}
\par
The measurement setup in PPMS allows for sample rotation in the $B$-field. In Fig.~\ref{fig:cross_section}, the geometry of the DC experiment is shown. A static magnetic field was applied in the z-direction. Fig.~\ref{fig:cross_section} presents a top view and side view of the two-finger InP HEMT where an angle of rotation $\theta$ is defined with respect to the magnetic field. $\theta$ could be varied between $0^\text{o}$ and $180^\text{o}$ degrees where $0^\text{o}$ ($180^\text{o}$) and $90^\text{o}$ corresponded to measurements of the InP HEMT oriented parallel and perpendicular to the magnetic field, respectively.

\begin{figure}[t]
    \centering
    \includegraphics[width=\columnwidth]{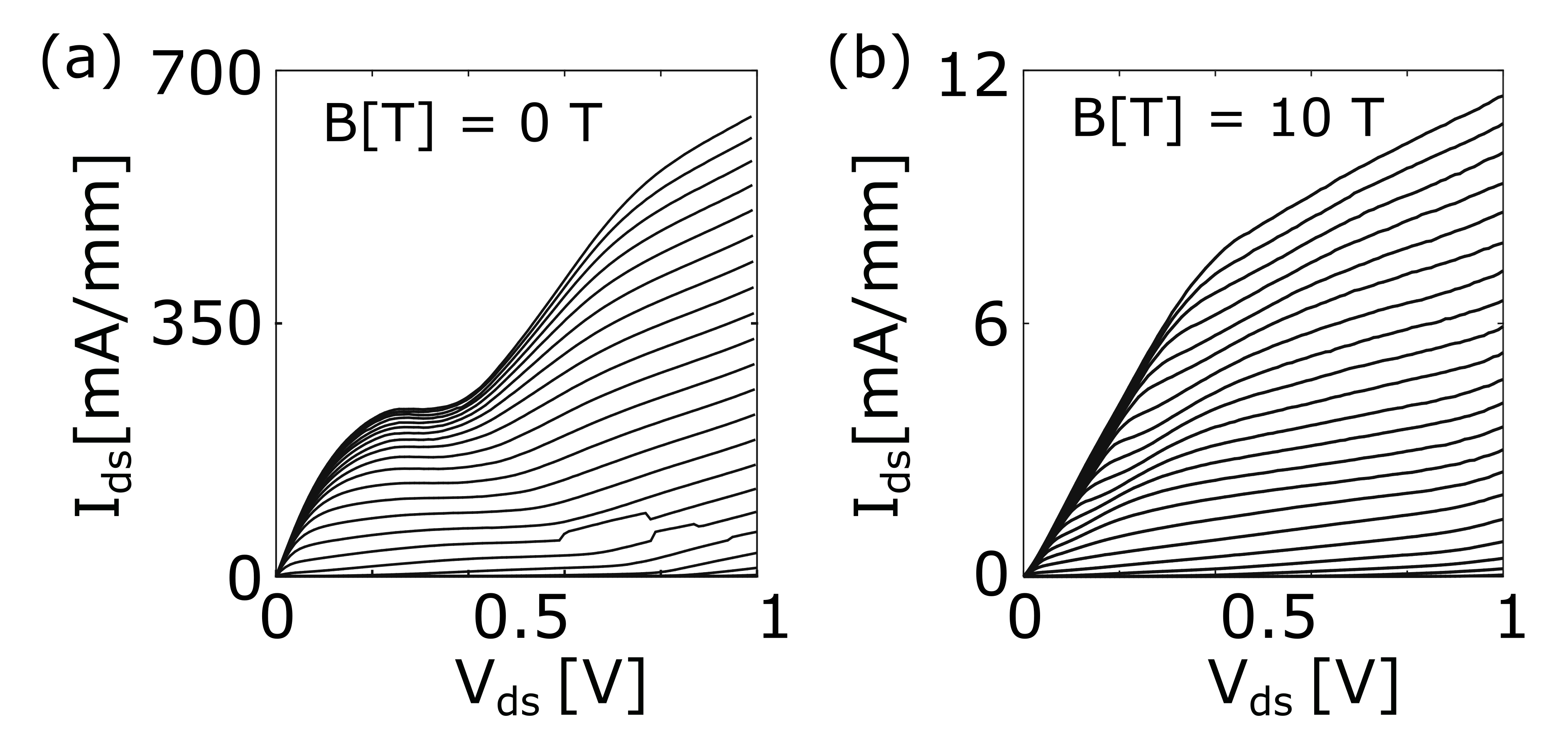}
    \caption{$I_{ds}$ versus $V_{ds}$ for the InP HEMT at 2 K oriented $\theta=90^\text{o}$ in a static magnetic field of (a) 0~T and (b) 10~T. $V_{gs}$ was varied from -0.4 to 0.4 V in steps on 0.1 V. $W_g=$2x50~$\mu$m and $Lg$=100~nm. Note difference in y-axis scale for (a) and (b).}
    \label{fig:Ids_vs_Vds}
\end{figure}

The cryogenic InP HEMT output current was first measured for $\theta=90^\text{o}$. Figure \ref{fig:Ids_vs_Vds} shows the source-drain current $I_{ds}$ versus source-drain voltage $V_{ds}$ for different gate-source voltage $V_{gs}$ under a magnetic field of 0~T and 10~T. The device was a 2x50~$\mu$m InP HEMT with $L_g$=100~nm. Figure \ref{fig:Ids_vs_Vds}~(a) illustrates a typical InP HEMT measured under cryogenic conditions in the absence of a magnetic field. The kink behavior at high current level around $V_{ds}$=0.4~V originates from electron traps in the interface layers of the heterostructure and is characteristic for these type of devices measured at low temperature.\cite{IEEERodilla15} Since the LC network in the sample holder did not provide a perfect 50~$\Omega$ impedance environment for the transistor, oscillations occurred for lower $I_{ds}$ showing up as small current jumps at $V_{gs}$~$<$~-0.2~V, which do not influence the interpretations in this study. In Fig.~\ref{fig:Ids_vs_Vds}~(b), the same measurement was repeated under a magnetic field of 10~T. It stands clear that $I_{ds}$ is extremely suppressed for $\theta=90^\text{o}$ when exposed to the magnetic field. Comparing Fig.~\ref{fig:Ids_vs_Vds}~(a) and (b), the maximum $I_{ds}$ ($V_{ds}$=1~V) was reduced by almost a factor of 100. The device still behaves as a transistor but with a much larger on-resistance and strongly reduced transconductance. 

\begin{figure}[t]
    \centering
    \includegraphics[width=\columnwidth]{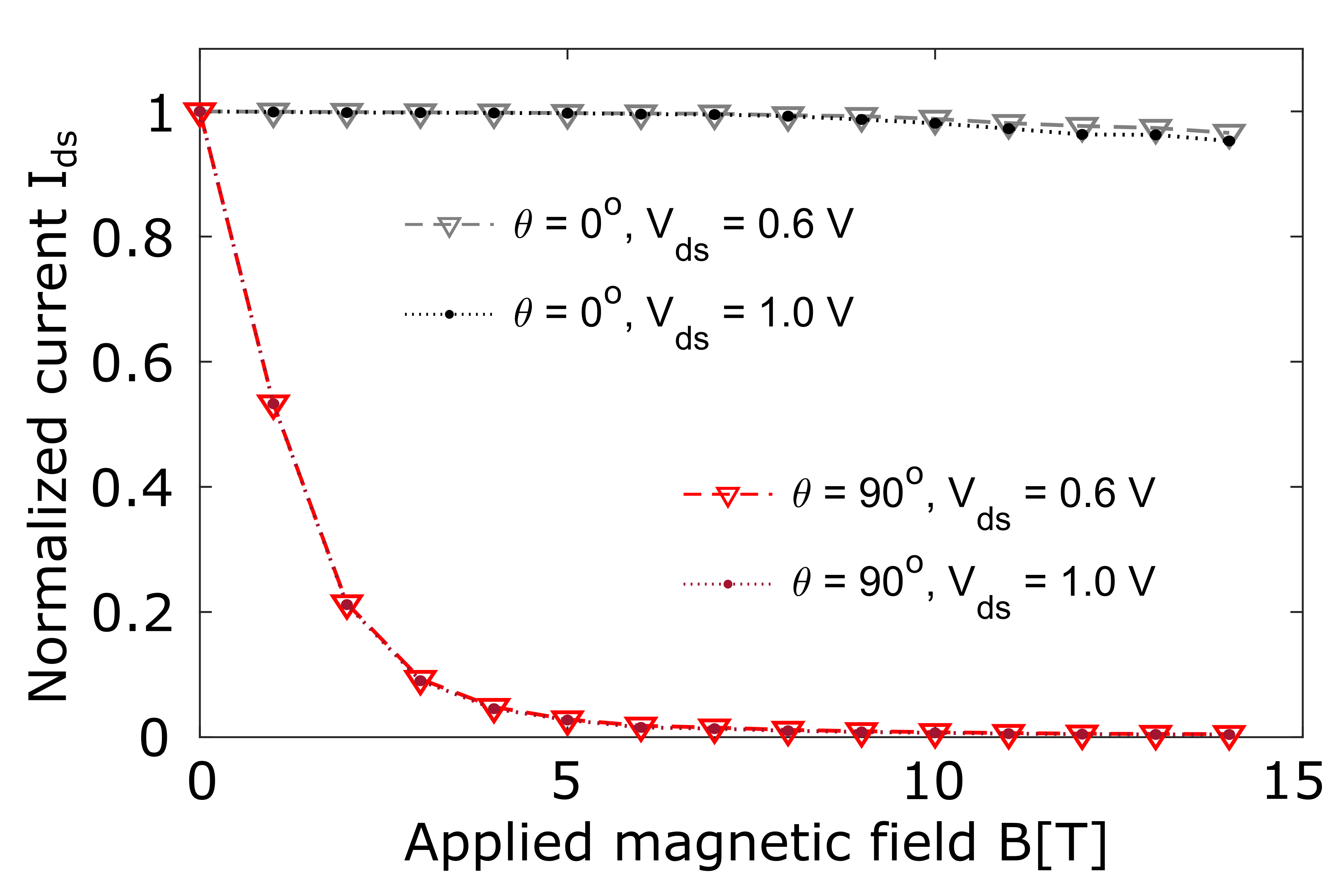}
    \caption{Normalized maximum $I_{ds}$ versus magnetic field ranging from 0 to 14~T in steps of 1 T with $\theta = 0^\text{o}$ (black) and $\theta = 90^\text{o}$ (red), at $V_{gs}= 0.4$~V. Values for $V_{ds}=0.6$~V (triangle) and 1.0~V (dot) are plotted. Device size: $W_g=2\text{x}50$~$\mu$m and $L_g=100$~nm, measured at 2~K.}
    \label{fig:Ids_vs_B}
\end{figure}

Measurements for the InP HEMT shown in Fig.~\ref{fig:Ids_vs_Vds} for $\theta=0^\text{o}$ and $\theta=90^\text{o}$ in a magnetic field up to 14~T are summarized in Fig.~\ref{fig:Ids_vs_B}. $I_{ds}$ was measured under saturation for two biases $V_{ds}$=0.4 and 0.6~V and normalized with respect to zero field. The strong suppression in output current for $\theta=90^\text{o}$ is visualized in Fig.~\ref{fig:Ids_vs_B}. As comparison, the effect in $I_{ds}$ from the magnetic field for $\theta=0^\text{o}$ is almost negligible (a minor reduction beyond 10~T may be due to a slight mis-orientation of the HEMT in the field). No difference with regard to $V_{ds}$ is noted in Fig.~\ref{fig:Ids_vs_B}.
\par
\begin{figure}[t]
    \centering
    \includegraphics[width=\columnwidth]{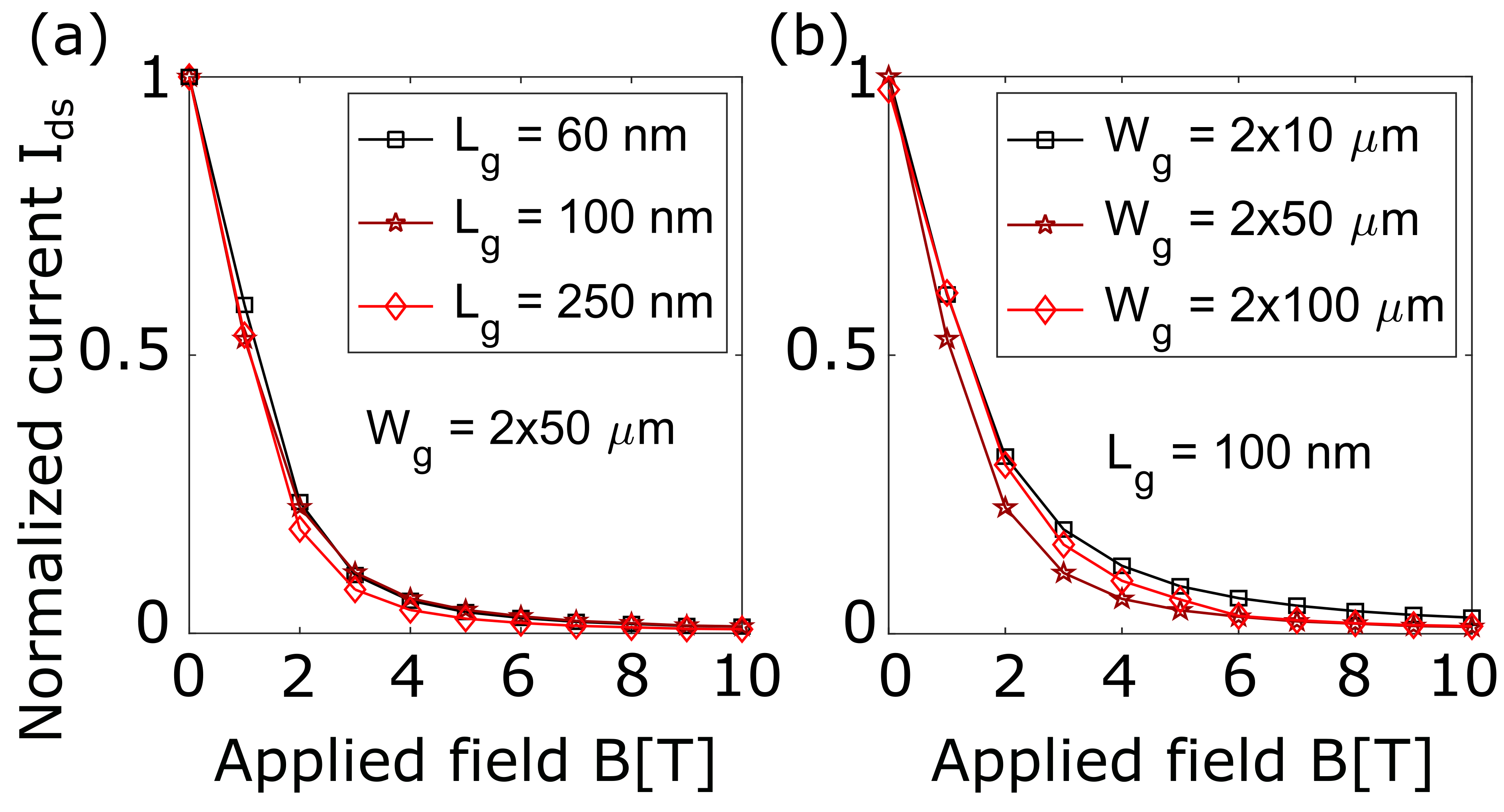}
    \caption{Normalized $I_{ds}$ versus magnetic field up to 10~T for InP HEMTs oriented $\theta=90^\text{o}$ with (a) $L_g=60$, 100 and 250~nm and (b) $W_g=2$x10, 2x50 and 2x100~$\mu$m. A fix $V_{gs}$ and $V_{ds}$ of 0.4~V were applied in an ambient temperature of 2~K.}
    \label{fig:variousLgWg}
\end{figure}

We also observe in Fig.~\ref{fig:Ids_vs_B} that for $\theta=90^\text{o}$, the normalized $I_{ds}$ varies as $\text{B}^2$.
Moreover, it was confirmed that this dependence was the same for a range of various device sizes $L_g$ (60, 100, 250~nm) and $W_g$ (2x10, 2x50, 2x100~$\mu$m), see Fig.~\ref{fig:variousLgWg}. In our experimental setup, the Hall effect (voltage) is negligible because of the device geometry with $W_g$ $>>$ $L_g$.\cite{gMR_Campbell11} The output current transport in the InP HEMT is therefore limited by gMR,\cite{gMR_JERVIS70} which occurs due to the effect of the Lorentz force on the 2DEG in devices where $W_g$ $>>$ $L_g$. gMR is expected to be large under the experimental circumstances here, \emph{i.e.} high channel mobility in the 2DEG and a strong magnetic field.

\par
\begin{figure}[t]
    \centering
    \includegraphics[width=\columnwidth]{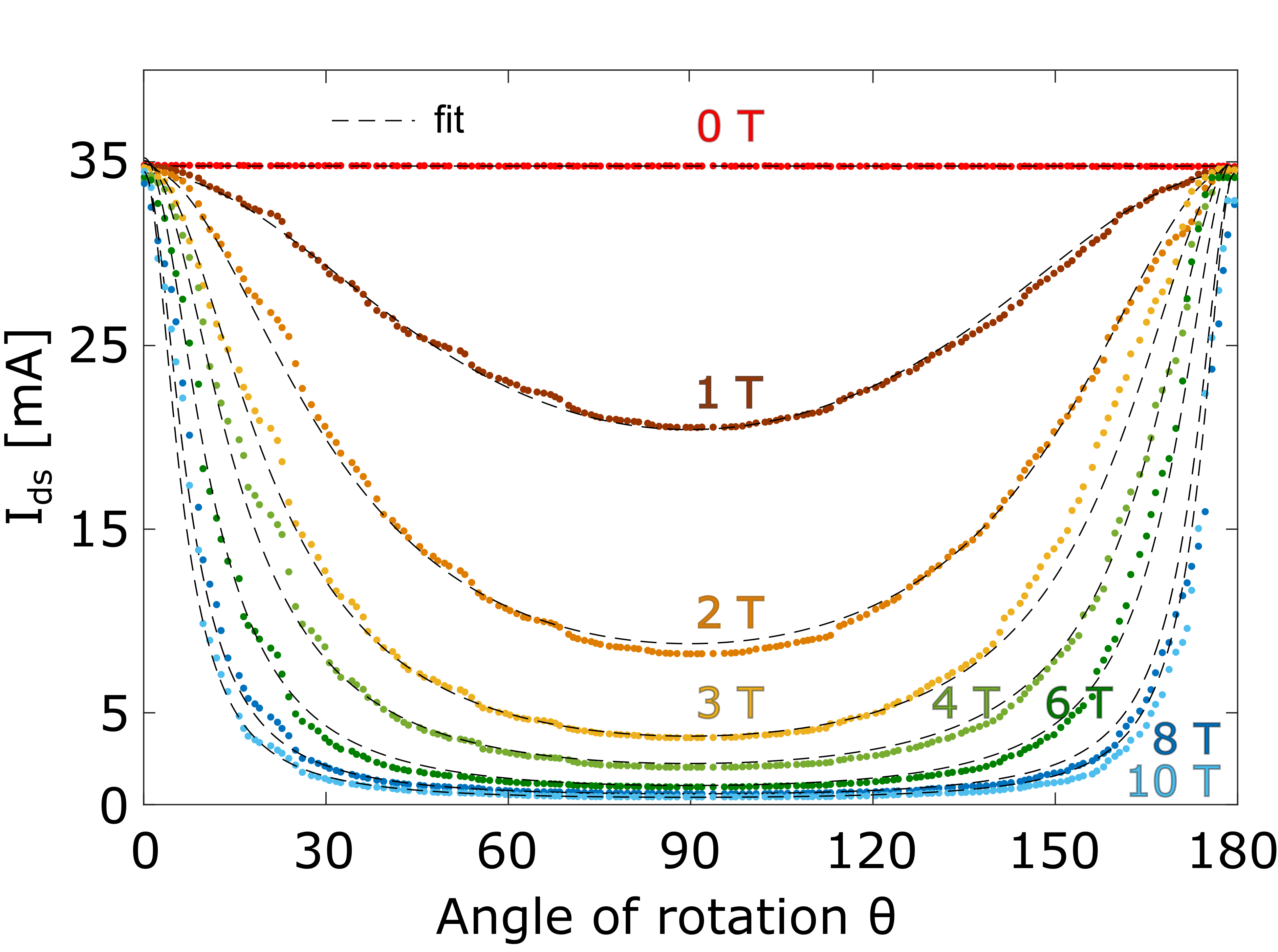}
    \caption{Rotation sweep ($\theta$ from 0 to $180^\text{o}$) showing $I_{ds}$ (absolute values) for various externally applied magnetic fields 0, 1, 2, 3, 4, 6, 8 and 10~T at a fix $V_{gs}$ and $V_{ds}$ of 0.4~V in an ambient temperature of 2~K. The dashed lines (black) are fitting of Eq.(\ref{eq:rot}) to the experimental data points (colored). Device size $W_g$=2x50~$\mu$m and $L_g=100$~nm.}
    \label{fig:rot_fit}
\end{figure}

The angular dependence of the cryogenic InP HEMT output current in a magnetic field was investigated by rotating the transistor in the magnetic field. $\theta$ was increased from $0^\text{o}$ to $180^\text{o}$ using a step size of $1^\text{o}$. In Fig.~\ref{fig:rot_fit}, $I_{ds}$ is plotted as a function of $\theta$ up to 10~T. It is noted that the $I_{ds}$ is clearly dependent on $\theta$ and that this dependence becomes larger with higher magnetic field. Irrespectively of the applied field strength, we observe no change in the output current when (the 2DEG channel of) the transistor has a 0$^\text{o}$ or 180$^\text{o}$ rotation. A significant reduction ($\sim20$~\%) in $I_{ds}$ for rotations as small as $15^\text{o}$ occurs at an applied field of 3~T. With increasing magnetic field, the alignment of the InP HEMT becomes even more crucial and the $I_{ds}$ is reduced by a minor tilt in $\theta$ of a few degrees.

 For a transistor layout with $W_g$~$>>$~$L_g$, the gMR in the channel is expected to vary as $1+\mu^2B^2$, where $\mu$ is the electron mobility in the channel. \cite{gMR_Campbell11}$^,$\cite{gMR_JERVIS70} Taking the angular dependence in Fig.~\ref{fig:cross_section} into account for a transistor output current in an electrical field subject to a Lorentz force leads to
\begin{equation}\label{eq:rot}
    I_{ds}(B,\theta) = \frac{V_{ds}}{R_0(1 + \mu^2B^2sin^2\theta)}
\end{equation}
where $R_0$ is a resistance term (including channel and access resistance contributions in the InP HEMT). Fitting the data to Eq.~\ref{eq:rot} gives $R_0=11$~$\Omega$ and $\mu=10,500$~cm$^2/$Vs, which is in excellent agreement for the full range of measured $\theta$ and B in Fig.~\ref{fig:rot_fit} reflecting the symmetrical angular dependence in $I_{ds}$ around $\theta=90^\text{o}$.



\begin{figure}[t]
    \centering
    \includegraphics[width=\columnwidth]{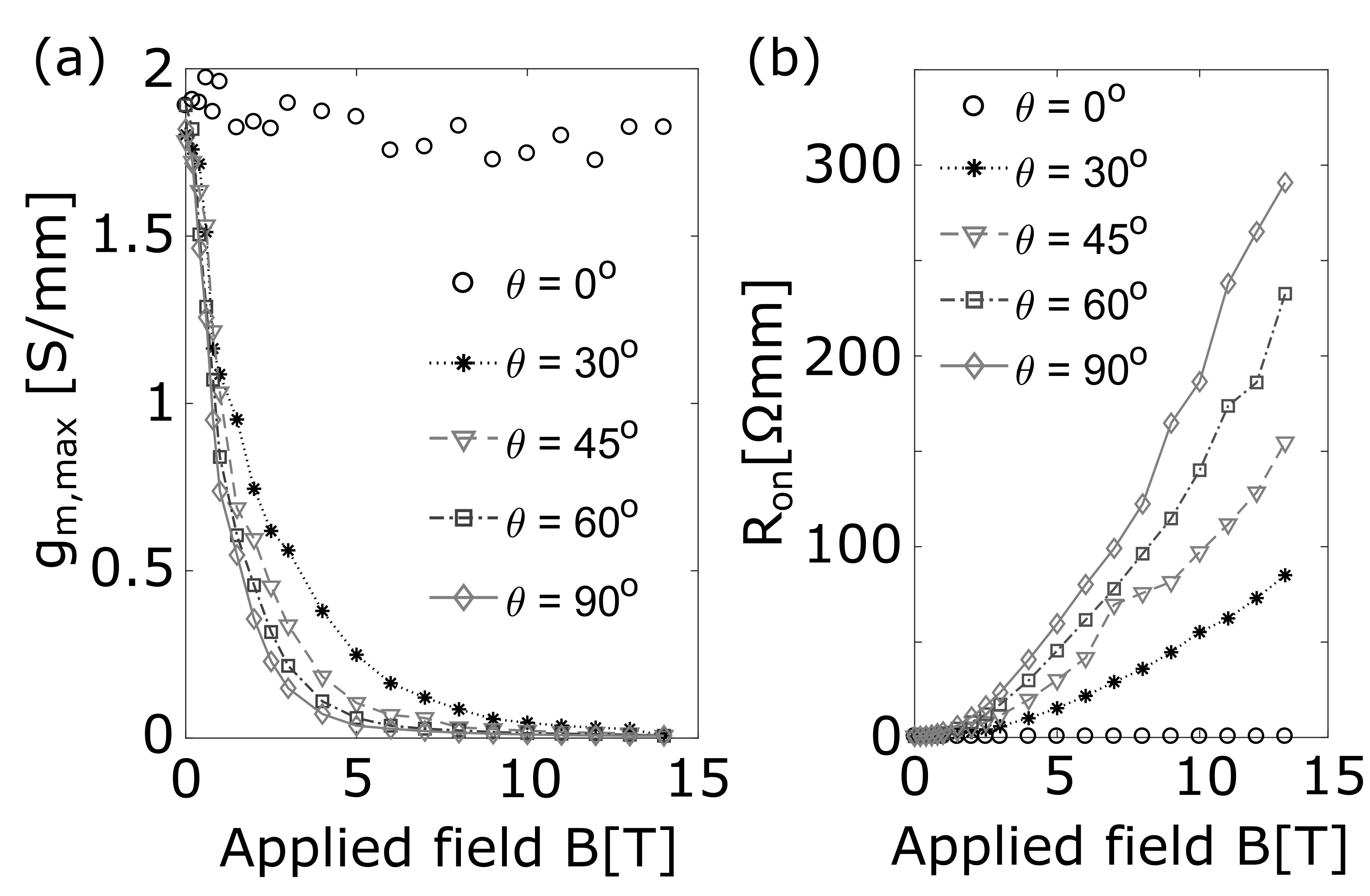}
    \caption{(a) $g_{m,max}$ and (b) $R_{on}$ as a function of applied magnetic field up to 14~T for $\theta = 0^\text{o},\ 30^\text{o},\ 45^\text{o},\ 60^\text{o}\ \text{and}\ 90^\text{o}$. Ambient temperature of 2~K and device size $W_g$ = 2x50~$\mu$m, $L_g$ = 100~nm. $V_{ds}$ = 1.0~V, $V_{gs}=0.4$~V.}
    \label{fig:gm_ron}
\end{figure}

The transconductance and on-resistance are two HEMT parameters fundamental for the design of a cryogenic LNA. The maximum transconductance g$_{m,max}$ at 2~K of the cryogenic InP HEMT is plotted in Fig.~\ref{fig:gm_ron}~(a) as a function of the applied magnetic field up to 14~T for various angles $\theta$. The g$_{m,max}$ is around 1.9~S/mm at zero field and strongly decreases as a function of magnetic field for $\theta=30^\text{o}$ and above. Already at 1~T, the g$_{m,max}$ is reduced with 40~\% (60~\%) for $\theta=30^\text{o}\ (90^\text{o})$. Field-effect transistor transconductance versus magnetic field for $\theta=90^\text{o}$ has been reported in Refs.~\cite{bradley} and~\cite{Bodart1998} for GaAs and silicon devices at cryogenic and room-temperature conditions, respectively. The dependence on magnetic field is much stronger for the InP HEMTs investigated in this study. In contrast to Ref.~\cite{bradley}, the g$_{m,max}$ here is directly calculated from measured I-V data for the transistor at cryogenic temperature in a magnetic field. Compared to Fig.~\ref{fig:gm_ron}~(a), Ref.~\cite{Bodart1998} shows a much weaker $g_m$(B) dependence which is probably related to the silicon-based transistor subject to study, a device normally not used in cryogenic LNAs. The on-resistance for the InP HEMT (at~2 K) as a function of $B$ for various $\theta$ is presented in Fig.~\ref{fig:gm_ron}~(b). For B=0~T, the $R_{on}$ is around 0.7~$\Omega$mm independent of $B$ for $\theta=0^\text{o}$. As expected from Figs.~\ref{fig:Ids_vs_Vds} and \ref{fig:Ids_vs_B}, $R_{on}$ increases rapidly with $B$ for $\theta=90^\text{o}$, two orders of magnitude at 5~T. This dependence is also strong at smaller angles as illustrated for $\theta=30^\text{o}$ in Fig.~\ref{fig:gm_ron}(b). $R_{on}$ is found to vary in a similar way as the denominator in Eq.~(\ref{eq:rot}): $R_{on}=R_0(1 + \mu^2B^2sin^2\theta)$. Fig.~\ref{fig:gm_ron} demonstrates that transistor parameters crucial for the design of a cryogenic LNA are highly sensitive to the alignment of the InP HEMT in a magnetic field at 2~K. This explains the strong degradation of the LNA gain and average noise temperature (measured for $\theta=90^\text{o}$) observed in Fig.~\ref{fig:gain_noise_LNA_90deg}.

In conclusion, we have investigated the angular dependence of the output current of cryogenic InP HEMTs in magnetic fields up to 14~T and found that it is greatly attenuated, not only at $\theta=90^\text{o}$, but also at small $\theta$. The physical reason is the very strong gMR occurring for $I_{ds}$ in the cryogenic InP HEMT. This was validated by an accurate fitting of experimental $I_{ds}$ data with an equation describing the gMR as a function of $B$ and $\theta$. Furthermore, we have shown the strong influence from $\theta$ for the transistor parameters $g_m$ and $R_{on}$  when the cryogenic InP HEMT is exposed to a magnetic field. As a result, the alignment of cryogenic InP HEMT LNAs with respect to a magnetic field is critical in sensitive microwave detection systems: even small deviation from $\theta = 0^\text{o}\ (180^\text{o})$ leads to significantly lower gain and higher average noise temperature.

\bigskip

This work was performed in GigaHertz Centre in a joint research project between Chalmers University of Technology, Low Noise Factory AB, Wasa Millimeter Wave AB, Omnisys Instruments AB and RISE Research Institutes of Sweden. We are grateful to Serguei Cherednichenko for valuable assistance in the noise measurements and Niklas Wadefalk for the LNA design.
\bigskip

\begin{thebibliography}{11}%
\makeatletter
\providecommand \@ifxundefined [1]{%
 \@ifx{#1\undefined}
}%
\providecommand \@ifnum [1]{%
 \ifnum #1\expandafter \@firstoftwo
 \else \expandafter \@secondoftwo
 \fi
}%
\providecommand \@ifx [1]{%
 \ifx #1\expandafter \@firstoftwo
 \else \expandafter \@secondoftwo
 \fi
}%
\providecommand \natexlab [1]{#1}%
\providecommand \enquote  [1]{``#1''}%
\providecommand \bibnamefont  [1]{#1}%
\providecommand \bibfnamefont [1]{#1}%
\providecommand \citenamefont [1]{#1}%
\providecommand \href@noop [0]{\@secondoftwo}%
\providecommand \href [0]{\begingroup \@sanitize@url \@href}%
\providecommand \@href[1]{\@@startlink{#1}\@@href}%
\providecommand \@@href[1]{\endgroup#1\@@endlink}%
\providecommand \@sanitize@url [0]{\catcode `\\12\catcode `\$12\catcode
  `\&12\catcode `\#12\catcode `\^12\catcode `\_12\catcode `\%12\relax}%
\providecommand \@@startlink[1]{}%
\providecommand \@@endlink[0]{}%
\providecommand \url  [0]{\begingroup\@sanitize@url \@url }%
\providecommand \@url [1]{\endgroup\@href {#1}{\urlprefix }}%
\providecommand \urlprefix  [0]{URL }%
\providecommand \Eprint [0]{\href }%
\providecommand \doibase [0]{http://dx.doi.org/}%
\providecommand \selectlanguage [0]{\@gobble}%
\providecommand \bibinfo  [0]{\@secondoftwo}%
\providecommand \bibfield  [0]{\@secondoftwo}%
\providecommand \translation [1]{[#1]}%
\providecommand \BibitemOpen [0]{}%
\providecommand \bibitemStop [0]{}%
\providecommand \bibitemNoStop [0]{.\EOS\space}%
\providecommand \EOS [0]{\spacefactor3000\relax}%
\providecommand \BibitemShut  [1]{\csname bibitem#1\endcsname}%
\let\auto@bib@innerbib\@empty
\bibitem [{\citenamefont {{Mathur}}, \citenamefont {{Knepper}},\ and\
  \citenamefont {{O'Connor}}(2008)}]{mathur2008}%
  \BibitemOpen
  \bibfield  {author} {\bibinfo {author} {\bibfnamefont {R.}~\bibnamefont
  {{Mathur}}}, \bibinfo {author} {\bibfnamefont {R.~W.}\ \bibnamefont
  {{Knepper}}}, \ and\ \bibinfo {author} {\bibfnamefont {P.~B.}\ \bibnamefont
  {{O'Connor}}},\ }\href {\doibase 10.1109/TASC.2008.2007272} {\bibfield
  {journal} {\bibinfo  {journal} {IEEE Transactions on Applied
  Superconductivity}\ }\textbf {\bibinfo {volume} {18}},\ \bibinfo {pages}
  {1781--1789} (\bibinfo {year} {2008})}\BibitemShut {NoStop}%
\bibitem [{\citenamefont {Hagmann}\ \emph {et~al.}(1990)\citenamefont
  {Hagmann}, \citenamefont {Sikivie}, \citenamefont {Sullivan},\ and\
  \citenamefont {Tanner}}]{PhysRevD.42.1297}%
  \BibitemOpen
  \bibfield  {author} {\bibinfo {author} {\bibfnamefont {C.}~\bibnamefont
  {Hagmann}}, \bibinfo {author} {\bibfnamefont {P.}~\bibnamefont {Sikivie}},
  \bibinfo {author} {\bibfnamefont {N.~S.}\ \bibnamefont {Sullivan}}, \ and\
  \bibinfo {author} {\bibfnamefont {D.~B.}\ \bibnamefont {Tanner}},\ }\href
  {\doibase 10.1103/PhysRevD.42.1297} {\bibfield  {journal} {\bibinfo
  {journal} {Phys. Rev. D}\ }\textbf {\bibinfo {volume} {42}},\ \bibinfo
  {pages} {1297--1300} (\bibinfo {year} {1990})}\BibitemShut {NoStop}%
\bibitem [{\citenamefont {{Brubaker}}(2018)}]{2018arXiv180100835B}%
  \BibitemOpen
  \bibfield  {author} {\bibinfo {author} {\bibfnamefont {B.~M.}\ \bibnamefont
  {{Brubaker}}},\ }\href@noop {} {\bibfield  {journal} {\bibinfo  {journal}
  {arXiv e-prints}\ ,\ \bibinfo {eid} {arXiv:1801.00835}} (\bibinfo {year}
  {2018})}\BibitemShut {NoStop}%
\bibitem [{\citenamefont {Johansen}\ \emph {et~al.}(2018)\citenamefont
  {Johansen}, \citenamefont {Sanchez-Heredia}, \citenamefont {Petersen},
  \citenamefont {Johansen}, \citenamefont {Zhurbenko},\ and\ \citenamefont
  {Ardenkjaer-Larsen}}]{Johansen18}%
  \BibitemOpen
  \bibfield  {author} {\bibinfo {author} {\bibfnamefont {D.~H.}\ \bibnamefont
  {Johansen}}, \bibinfo {author} {\bibfnamefont {J.~D.}\ \bibnamefont
  {Sanchez-Heredia}}, \bibinfo {author} {\bibfnamefont {J.~R.}\ \bibnamefont
  {Petersen}}, \bibinfo {author} {\bibfnamefont {T.~K.}\ \bibnamefont
  {Johansen}}, \bibinfo {author} {\bibfnamefont {V.}~\bibnamefont {Zhurbenko}},
  \ and\ \bibinfo {author} {\bibfnamefont {J.~H.}\ \bibnamefont
  {Ardenkjaer-Larsen}},\ }\href {\doibase 10.1109/TBCAS.2017.2776256}
  {\bibfield  {journal} {\bibinfo  {journal} {IEEE Transactions on Biomedical
  Circuits and Systems}\ }\textbf {\bibinfo {volume} {12}},\ \bibinfo {pages}
  {202--210} (\bibinfo {year} {2018})}\BibitemShut {NoStop}%
\bibitem [{\citenamefont {Daw}\ and\ \citenamefont {Bradley}(1997)}]{bradley}%
  \BibitemOpen
  \bibfield  {author} {\bibinfo {author} {\bibfnamefont {E.}~\bibnamefont
  {Daw}}\ and\ \bibinfo {author} {\bibfnamefont {R.~F.}\ \bibnamefont
  {Bradley}},\ }\href@noop {} {\bibfield  {journal} {\bibinfo  {journal}
  {Journal of Applied Physics}\ }\textbf {\bibinfo {volume} {82(4)}},\ \bibinfo
  {pages} {1925--1929} (\bibinfo {year} {1997})}\BibitemShut {NoStop}%
\bibitem [{\citenamefont {{Schleeh}}\ \emph {et~al.}(2013)\citenamefont
  {{Schleeh}}, \citenamefont {{Wadefalk}}, \citenamefont {{Nilsson}},
  \citenamefont {{Starski}},\ and\ \citenamefont
  {{Grahn}}}]{mmic_schleeh13TMTT}%
  \BibitemOpen
  \bibfield  {author} {\bibinfo {author} {\bibfnamefont {J.}~\bibnamefont
  {{Schleeh}}}, \bibinfo {author} {\bibfnamefont {N.}~\bibnamefont
  {{Wadefalk}}}, \bibinfo {author} {\bibfnamefont {P.}~\bibnamefont
  {{Nilsson}}}, \bibinfo {author} {\bibfnamefont {J.~P.}\ \bibnamefont
  {{Starski}}}, \ and\ \bibinfo {author} {\bibfnamefont {J.}~\bibnamefont
  {{Grahn}}},\ }\href {\doibase 10.1109/TMTT.2012.2235856} {\bibfield
  {journal} {\bibinfo  {journal} {IEEE Transactions on Microwave Theory and
  Techniques}\ }\textbf {\bibinfo {volume} {61}},\ \bibinfo {pages} {871--877}
  (\bibinfo {year} {2013})}\BibitemShut {NoStop}%
\bibitem [{\citenamefont {{Schleeh}}\ \emph {et~al.}(2012)\citenamefont
  {{Schleeh}}, \citenamefont {{Alestig}}, \citenamefont {{Halonen}},
  \citenamefont {{Malmros}}, \citenamefont {{Nilsson}}, \citenamefont
  {{Nilsson}}, \citenamefont {{Starski}}, \citenamefont {{Wadefalk}},
  \citenamefont {{Zirath}},\ and\ \citenamefont {{Grahn}}}]{hemt_schleeh12EDL}%
  \BibitemOpen
  \bibfield  {author} {\bibinfo {author} {\bibfnamefont {J.}~\bibnamefont
  {{Schleeh}}}, \bibinfo {author} {\bibfnamefont {G.}~\bibnamefont
  {{Alestig}}}, \bibinfo {author} {\bibfnamefont {J.}~\bibnamefont
  {{Halonen}}}, \bibinfo {author} {\bibfnamefont {A.}~\bibnamefont
  {{Malmros}}}, \bibinfo {author} {\bibfnamefont {B.}~\bibnamefont
  {{Nilsson}}}, \bibinfo {author} {\bibfnamefont {P.~A.}\ \bibnamefont
  {{Nilsson}}}, \bibinfo {author} {\bibfnamefont {J.~P.}\ \bibnamefont
  {{Starski}}}, \bibinfo {author} {\bibfnamefont {N.}~\bibnamefont
  {{Wadefalk}}}, \bibinfo {author} {\bibfnamefont {H.}~\bibnamefont
  {{Zirath}}}, \ and\ \bibinfo {author} {\bibfnamefont {J.}~\bibnamefont
  {{Grahn}}},\ }\href {\doibase 10.1109/LED.2012.2187422} {\bibfield  {journal}
  {\bibinfo  {journal} {IEEE Electron Device Letters}\ }\textbf {\bibinfo
  {volume} {33}},\ \bibinfo {pages} {664--666} (\bibinfo {year}
  {2012})}\BibitemShut {NoStop}%
\bibitem [{\citenamefont {Rodilla}\ \emph {et~al.}(2015)\citenamefont
  {Rodilla}, \citenamefont {Schleeh}, \citenamefont {Nilsson},\ and\
  \citenamefont {Grahn}}]{IEEERodilla15}%
  \BibitemOpen
  \bibfield  {author} {\bibinfo {author} {\bibfnamefont {H.}~\bibnamefont
  {Rodilla}}, \bibinfo {author} {\bibfnamefont {J.}~\bibnamefont {Schleeh}},
  \bibinfo {author} {\bibfnamefont {P.}~\bibnamefont {Nilsson}}, \ and\
  \bibinfo {author} {\bibfnamefont {J.}~\bibnamefont {Grahn}},\ }\href@noop {}
  {\bibfield  {journal} {\bibinfo  {journal} {IEEE Trans. Electron Devices}\
  }\textbf {\bibinfo {volume} {62(2)}},\ \bibinfo {pages} {532--537} (\bibinfo
  {year} {2015})}\BibitemShut {NoStop}%
\bibitem [{\citenamefont {{Campbell}}\ \emph {et~al.}(2011)\citenamefont
  {{Campbell}}, \citenamefont {{Cheung}}, \citenamefont {{Yu}}, \citenamefont
  {{Suehle}}, \citenamefont {{Oates}},\ and\ \citenamefont
  {{Sheng}}}]{gMR_Campbell11}%
  \BibitemOpen
  \bibfield  {author} {\bibinfo {author} {\bibfnamefont {J.~P.}\ \bibnamefont
  {{Campbell}}}, \bibinfo {author} {\bibfnamefont {K.~P.}\ \bibnamefont
  {{Cheung}}}, \bibinfo {author} {\bibfnamefont {L.~C.}\ \bibnamefont {{Yu}}},
  \bibinfo {author} {\bibfnamefont {J.~S.}\ \bibnamefont {{Suehle}}}, \bibinfo
  {author} {\bibfnamefont {A.}~\bibnamefont {{Oates}}}, \ and\ \bibinfo
  {author} {\bibfnamefont {K.}~\bibnamefont {{Sheng}}},\ }\href {\doibase
  10.1109/LED.2010.2086044} {\bibfield  {journal} {\bibinfo  {journal} {IEEE
  Electron Device Letters}\ }\textbf {\bibinfo {volume} {32}},\ \bibinfo
  {pages} {75--77} (\bibinfo {year} {2011})}\BibitemShut {NoStop}%
\bibitem [{\citenamefont {Jervis}\ and\ \citenamefont
  {Johnson}(1970)}]{gMR_JERVIS70}%
  \BibitemOpen
  \bibfield  {author} {\bibinfo {author} {\bibfnamefont {T.}~\bibnamefont
  {Jervis}}\ and\ \bibinfo {author} {\bibfnamefont {E.}~\bibnamefont
  {Johnson}},\ }\href {\doibase https://doi.org/10.1016/0038-1101(70)90049-3}
  {\bibfield  {journal} {\bibinfo  {journal} {Solid-State Electronics}\
  }\textbf {\bibinfo {volume} {13}},\ \bibinfo {pages} {181 -- 189} (\bibinfo
  {year} {1970})}\BibitemShut {NoStop}%
\bibitem [{\citenamefont {Bodart}\ \emph {et~al.}(1998)\citenamefont {Bodart},
  \citenamefont {Garcia}, \citenamefont {Phelps}, \citenamefont {Sullivan},
  \citenamefont {Moulton},\ and\ \citenamefont {Kuhns}}]{Bodart1998}%
  \BibitemOpen
  \bibfield  {author} {\bibinfo {author} {\bibfnamefont {J.~R.}\ \bibnamefont
  {Bodart}}, \bibinfo {author} {\bibfnamefont {B.~M.}\ \bibnamefont {Garcia}},
  \bibinfo {author} {\bibfnamefont {L.}~\bibnamefont {Phelps}}, \bibinfo
  {author} {\bibfnamefont {N.~S.}\ \bibnamefont {Sullivan}}, \bibinfo {author}
  {\bibfnamefont {W.~G.}\ \bibnamefont {Moulton}}, \ and\ \bibinfo {author}
  {\bibfnamefont {P.}~\bibnamefont {Kuhns}},\ }\href {\doibase
  10.1063/1.1148517} {\bibfield  {journal} {\bibinfo  {journal} {Review of
  Scientific Instruments}\ }\textbf {\bibinfo {volume} {69}},\ \bibinfo {pages}
  {319--320} (\bibinfo {year} {1998})}\BibitemShut {NoStop}%
\end{thebibliography}

%

\end{document}